\documentclass[reprint,prd,twocolumn,superscriptaddress,showpacs,floatfix]{revtex4}

\usepackage{amssymb}
\usepackage{graphicx}
\usepackage{epsfig}
\newcommand{\ba}{\begin{eqnarray}}
\newcommand{\ea}{\end{eqnarray}}
\newcommand{\be}{\begin{equation}}
\newcommand{\ee}{\end{equation}}
\newcommand{\beq}{\begin{equation}} 
\newcommand{\eeq}{\end{equation}}   
\newcommand{\bea}{\begin{eqnarray}} 
\newcommand{\eea}{\end{eqnarray}}
\def\Li2{\hbox{Li}_2}

\begin{document}


\title{Nucleon form factors and final state radiative corrections to 
  $e^+e^- \to \bar p p \gamma$}

\thanks{Work 
supported in part by
the Polish National Science Centre, grant number DEC-2012/07/B/ST2/03867 and
German Research Foundation DFG under
Contract No. Collaborative Research Center CRC-1044. }


\author{Henryk Czy\.z}
\affiliation{Institute of Physics, University of Silesia,
PL-40007 Katowice, Poland.}
\author{Johann H. K\"uhn}
\affiliation{Institut f\"ur Theoretische Teilchenphysik,
Karlsruhe Institute of Technology, D-76128 Karlsruhe, Germany.}
\author{Szymon Tracz}
\affiliation{Institute of Physics, University of Silesia,
PL-40007 Katowice, Poland.}


\date{\today}

\begin{abstract}
New parametrisation for the electric and the magnetic form factors of 
proton and neutron are presented. The proton form factors describe well
 the recent measurements by BaBar collaboration and earlier ones
 of the ratio of the form factors in space-like region.
  The neutron form factors
 are consistent with earlier measurements of neutron pair production
 and ratio of the form factors in the space-like region.
   These form factors are
  implemented into the generator PHOKHARA, which simulates the 
 reactions $e^+e^-\to \bar p p \gamma$ and $e^+e^-\to \bar n n\gamma$.
  The influence of final state radiation is investigated.

\end{abstract}


\maketitle

\newcommand{\Eq}[1]{Eq.(\ref{#1})} 

\section{\label{sec1}Introduction}

  Nucleon form factors both in the space-like and in the time-like 
 regions have attracted theoretical as well as experimental attention
 from the early times of particle physics 
  \cite{Stern,Rosenbluth:1950yq,Mcallister:1956ng,Hofstadter:1956qs}. 
 Experiments for electron-proton scattering give access to the 
 space-like region and detailed analyses of angular distributions 
 without \cite{Rosenbluth:1950yq} and with \cite{Akhiezer:1968ek}
 polarised beam and target allow to separate the magnetic and 
 electric form factors. The cross section for proton-antiproton production at 
 electron-positron colliders (or the inverse reaction $p\bar p \to e^+e^-$)
 gives access to a specific combination of form factors in the time-like 
 region. The additional analysis of angular distributions allows 
 to separate magnetic and electric form factors. Recently discrepancies 
  have been found between  the ratio of magnetic and electric 
 form factors extracted in the space-like region  using the Rosenbluth method
 or, alternatively, using polarised beam and/or target 
 (see \cite{Arrington:2003df} for a review).  
 The mentioned inconsistency was to large extent explained theoretically
  by enhanced contributions from two photon exchange
   (see \cite{Carlson:2007sp}
  for a review).
 Nevertheless this has triggered  a new
  experiment  (OLYMPUS) \cite{Milner:2013daa} to measure
 both electron and positron scattering
 off protons and designed to  resolve this issue experimentally.
  Form factors in the
 space- and time-like regions are connected via analyticity. Thus
  measurements in $e^+e^-$ collisions may help to resolve this issue.
 Even more important, these measurements are ideally suited 
 to search  for baryon-antibaryon resonances close to production threshold
 as well as at higher energies. Last but not least, the high energy
 behaviour has been predicted in the frame work of perturbative QCD
 \cite{Lepage:1980fj}, predictions that could be checked at high energy.
  For recent reviews on nucleon electromagnetic form factors
  we refer the reader to \cite{Perdrisat:2006hj} and \cite{Denig:2012by}.

 There are three reactions that are currently used for measurements
 in the time-like region: $e^+e^-\to p \bar p$, $p \bar p\to e^+e^-$
 and the radiative return reaction $e^+e^-\to p \bar p \gamma$.
 Given sufficiently large luminosity, the third reaction allows
 to measure the form factors in principle from threshold up to the
 collider energy. The radiative return has been employed successfully
 by the BaBar experiment \cite{Lees:2013ebn}, which has measured the production
  rate and the   ratio of the form factors with a precision of 7\% and  11\%
 respectively. In view of this improvement we present a parametrisation 
 of the nucleon form factors, which is based on generalised vector
 dominance, similar to that from \cite{Czyz:2010hj} for the case
 of pion and kaon pair production. Furthermore we consider in detail
 the impact of final state radiation to this measurement.
 In our simplified model we treat real radiation similar to the radiation
 from a point-like particle. As far as the virtual corrections are
 concerned we include the Coulomb enhancement factor, important 
 close to threshold and an infrared subtraction term to compensate 
 for soft real radiation.  These ingredients are
 implemented into Monte Carlo event generator PHOKHARA, version 9.1
 and the effect of these modifications is studied in detail.
 Additionally, the same modifications are added to the description 
 of proton-antiproton pair production in the scanning mode
  ($e^+e^-\to p \bar p$)   in  PHOKHARA version 9.1.

\section{\label{sec2} Nucleon form factors}

 In the first implementation \cite{Czyz:2004ua} of nucleon pair production
through the radiative return
 ($e^+e^- \to \bar N N \gamma$) into the event generator PHOKHARA
 a model of the nucleon form factors was used, which had been
 taken from \cite{Iachello:1972nu}.
   To accommodate
 the experimental data, which are available now, we propose a new, 
 improved model 
 for the  Dirac and Pauli nucleon form factors

 \begin{eqnarray}
F_{1,2}^p&=&F_{1,2}^s+F_{1,2}^v,\label{izprot}\\
F_{1,2}^n&=&F_{1,2}^s-F_{1,2}^v,\label{izneut}\,
\end{eqnarray}

which enter the electromagnetic current
\begin{equation}
J_{\mu}=-ie\bar{v}(p_2)\left(F_1^N(Q^2)\gamma_{\mu}-\frac{F_2^N(Q^2)}{4m_N}[\gamma_{\mu},\not{Q}]\right)u(p_1);
\label{pradh}
\end{equation} 
  where $Q=p_1+p_2$. The indices $s$ and $v$ refer to isospin zero
  and one respectively. 

 We use the following ansatz
 
\begin{equation}
 F_1^s=\frac{1}{2}\frac{\sum_{n=0}^4 c_n^1 BW_{\omega_n}(s)}{\sum_{n=0}^4 c_n^1},
\end{equation}
\begin{equation}
 F_1^v=\frac{1}{2}\frac{\sum_{n=0}^4 c_n^2 BW_{\rho_n}(s)}{\sum_{n=0}^4 c_n^2},
\end{equation}
\begin{equation}
F_2^s=-\frac{1}{2}b\frac{\sum_{n=0}^4 c_n^3 BW_{\omega_n}(s)}{\sum_{n=0}^4 c_n^3},
\end{equation}
\begin{equation}
 F_2^v=\frac{1}{2}a\frac{\sum_{n=0}^4 c_n^4 BW_{\rho_n}(s)}{\sum_{n=0}^4 c_n^4},
\end{equation}
where $c_0^i =1$ for $i=1,2,3,4$.
 Following the Zweig rule we neglect the $\phi$ contributions.
 The Breit - Wigner function is defined as:
\begin{equation}
BW_i(Q^2)=\frac{m_i^2}{m_i^2-Q^2-im_i\Gamma_i\theta(Q^2)} .
\end{equation}
  The  step function $\theta(Q^2)$  sets the mesons 
 widths to zero for space-like $Q^2$. 
 Above the proton-anti-proton production
 threshold we use constant meson widths. 
The normalisation of electric and magnetic form factors in the limit of $s=0$ to electric charges and magnetic moments of nucleons fixes the 
 parameters $a=\mu_p-\mu_n-1$ and $b=-\mu_p-\mu_n+1$, where $\mu_p(\mu_n)$
 are the magnetic moments of proton (neutron).  
 We impose the asymptotic (large $Q^2$) behaviour of the form factors
 predicted in perturbative $QCD$ \cite{Lepage:1980fj} 
\begin{equation}
F_1\sim\frac{1}{(Q^2)^2}, \hspace{5mm}F_2\sim \frac{1}{(Q^2)^3},
\label{asympt}
\end{equation}

\noindent
which leaves six independent complex parameters to be determined
by experimental data.
Below we rewrite them using real parameters
  $c_i^j = c_i^{jR}+ic_i^{jI}\theta(Q^2)$.

 The asymptotic behaviour, Eq.(\ref{asympt}), is enforced by choosing

\begin{eqnarray}
c_4^1&=&-\frac{1}{m_{\omega_4}^2}\sum_{n=0}^3 m_{\omega_n}^2c^1_n\\
c_4^2&=&-\frac{1}{m_{\rho_4}^2}\sum_{n=0}^3 m_{\rho_n}^2c^2_n\\
c_3^3 &=&\frac{\sum_{n=0}^2 m^2_{\omega_n}c_n^3(m_{\omega_n}^2-m_{\omega_4}^2+i(m_{\omega_4}\Gamma_{\omega_4}-m_{\omega_n}\Gamma_{\omega_n}))}{m_{\omega_3}^2 (m_{\omega_4}^2-m_{\omega_3}^2+i(m_{\omega_3}\Gamma_{\omega_3}-m_{\omega_4}\Gamma_{\omega_4}))}, \nonumber \\
c_4^3&=&-\frac{1}{m_{\omega_4}^2}\sum_{n=0}^3 m_{\omega_n}^2c^3_n\\
c_3^4&=&\frac{\sum_{n=0}^2 m^2_{\rho_n}c_n^4(m_{\rho_n}^2-m_{\rho_4}^2+i(m_{\rho_4}\Gamma_{\rho_4}-m_{\rho_n}\Gamma_{\rho_n}))}{m_{\rho_3}^2 (m_{\rho_4}^2-m_{\rho_3}^2+i(m_{\rho_3}\Gamma_{\rho_3}-m_{\rho_4}\Gamma_{\rho_4}))}, \nonumber\\
c_4^4&=&-\frac{1}{m_{\rho_4}^2}\sum_{n=0}^3 m_{\rho_n}^2c^4_n.\\
\end{eqnarray}

 The Dirac and  Pauli form factors for each 
 nucleon $N$ are related to the electric $G_E^N$ and magnetic
 $G_M^N$ through ($\tau = Q^2/4m_N^2$)

\begin{eqnarray}
G_E^N&=&F_1^N+\tau F_2^N,\label{ge}\\
G_M^N&=&F_1^N+F_2^N. \label{gm}
\end{eqnarray}

\begin{table}[h]
\begin{center}
\vskip0.3cm
\begin{tabular}{|c|c|c|c|c|c|}
\hline
Experiment &  nep  & $\chi^2$ & Experiment &  nep  & $\chi^2$  \\
\hline
BaBar cs \cite{Lees:2013ebn} &38& 30 & BaBar r \cite{Lees:2013ebn}&6&0.6 \\
PS170$_1$ cs \cite{psrat}&8&109&PS170 r \cite{psrat}&5&16 \\
PS170$_2$ cs \cite{ps170_1}&4&4&PS170$_3$ cs \cite{Bardin:1991bz}& 4&52\\
E760$_1$ cs \cite{E760}&3&0.5&E835$_1$ cs \cite{E835_1}&5& 1 \\
E835$_2$ cs \cite{E835_2}&2&0.03&DM2 cs \cite{dm2,dm22}&7& 26\\
BES cs \cite{bes}&8& 10&CLEO cs \cite{cleo}&1&0.4\\
FENICE cs \cite{fenice}&5& 5&DM1 cs \cite{dm1}&4&0.7\\
JLab 05 r \cite{jlab}&10&16& JLab 02 r \cite{jlab2002}&4&1\\
JLab 01 r \cite{jlab3}&13&10&JLab 10 r \cite{a23}&3&6 \\
MAMI 01 r \cite{mami}&3&2&JLab 03 r \cite{E}&3&6 \\
BLAST 08 r \cite{blast}&4&6&FENICE cs \cite{fenice}&4&0.6 \\
 & &  &SLAC cs \cite{Andivahis:1994rq}& 32& 27\\
\hline
\end{tabular}
\caption{{\it Values of the chi-squared distribution for particular experiments;  nep- number of experimental points; cs - cross section;
r- ratio of the electric and magnetic form factors. The PS170 and DM2 data were excluded from this fit - see text for details. }}
\label{tab:fit1}
\end{center}
\end{table} 

\begin{table}[h]
\begin{center}
\vskip0.3cm
\begin{tabular}{|c|c|c|c|c|c|c|c|}
\hline
$c_1^{1R}$   & -0.45(1)  & $c_1^{1I}$   & -0.54(2) &
$c_2^{1R}$   &-0.27(1) & $c_2^{1I}$   & 0.18(1) \\
$c_3^{1R}$   &  0.42(2)& $c_3^{1I}$   & 0.37(2)  &
$c_1^{2R}$   &  -0.12(1) & $c_1^{2I}$   &  -3.06(2)\\
$c_2^{2R}$   &  0.16(1) & $c_2^{2I}$  &  2.53(1)&
$c_3^{2R}$   &  -0.32(1) & $c_3^{2I}$   &  -0.17(1)\\
$c_1^{3R}$   &  -8.03(5) & $c_1^{3I}$  &  3.28(2)&
$c_2^{3R}$   &  10.6(1) & $c_2^{3I}$   &  0.2(3)\\
$c_1^{4R}$   & -0.845(1)  & $c_1^{4I}$  &  0.364(1)&
$c_2^{4R}$   &  0.427(1) & $c_2^{4I}$   &  -0.305(1)\\
\hline
\end{tabular}
\caption{{\it Parameters of the nucleons form factor 
 for the fit, where the PS170 and DM2 data were excluded- see text for details.}}
\label{tab:pion}
\end{center}
\end{table}

\begin{figure}[h]

\begin{center}
\includegraphics[width=8.cm,height=6.cm]{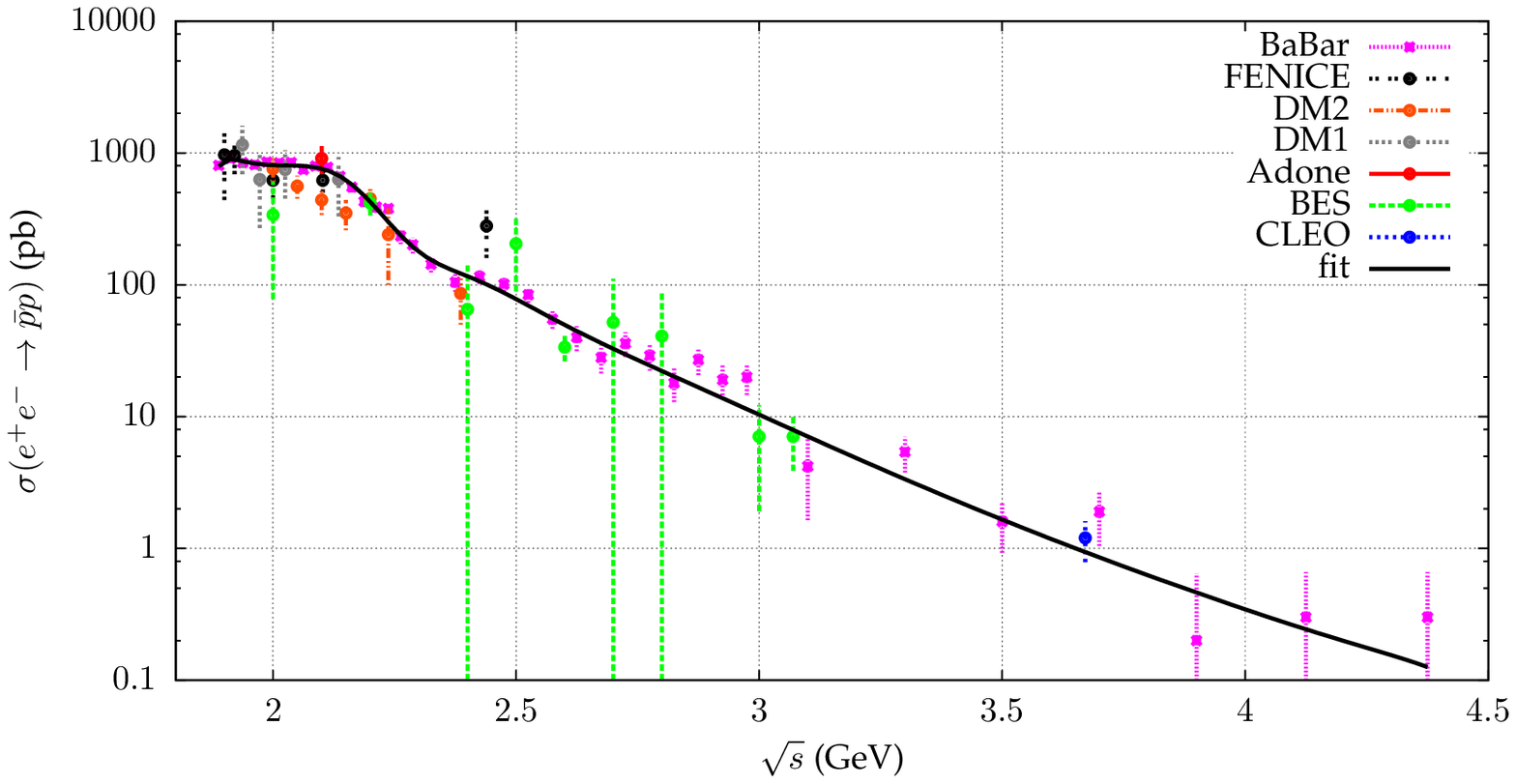}
\caption{(color online)
The experimental data compared to the model fits results.
 \label{pipi}
}
\end{center}
\end{figure}

\begin{figure}[ht]

\begin{center}
\includegraphics[width=8.3cm,height=6.cm]{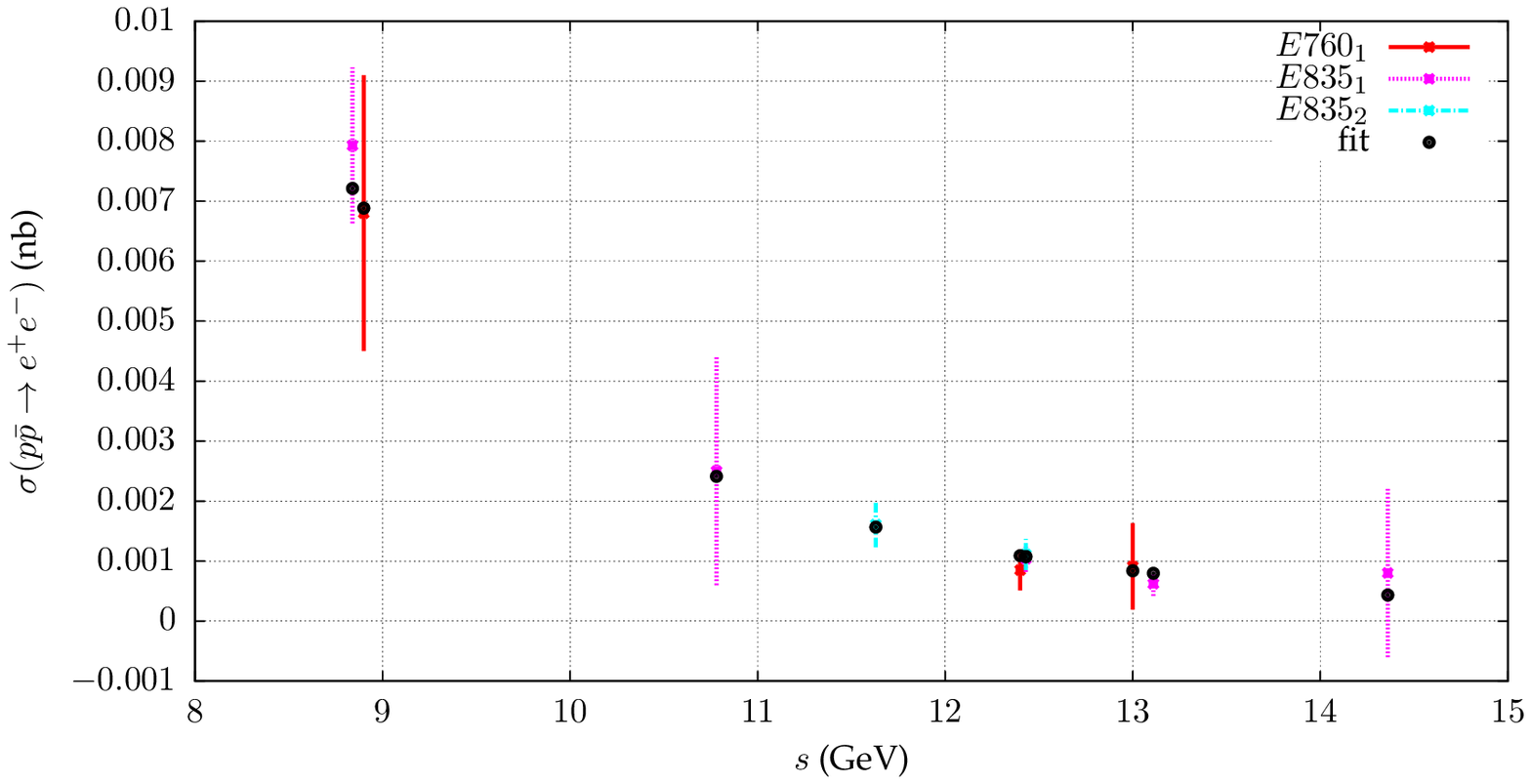}
\includegraphics[width=8.cm,height=6.cm]{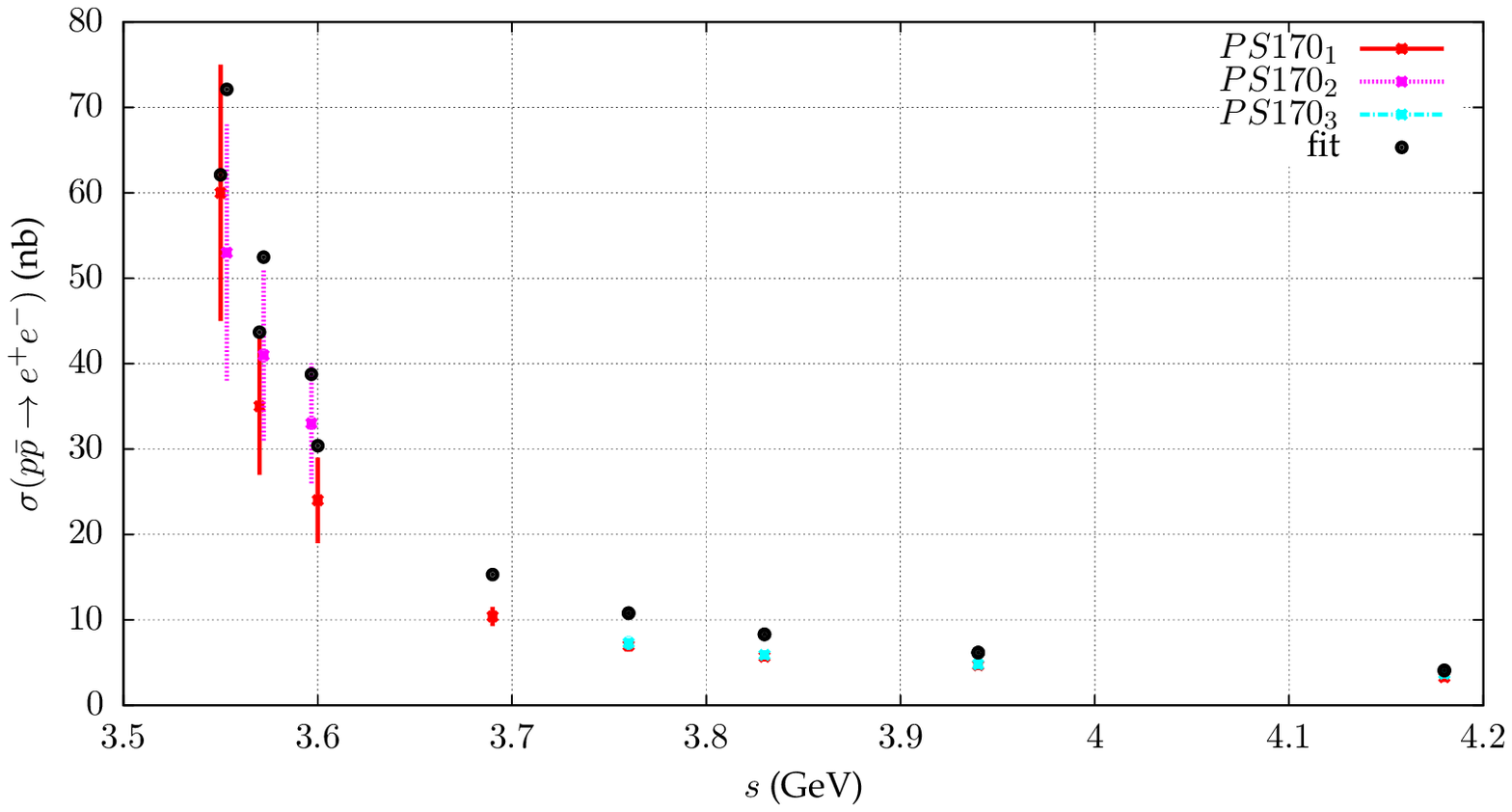}
\caption{(color online)
The experimental data compared to the model fits results.
 \label{pipi3}
}
\end{center}
 \vspace{0.5 cm}

\end{figure}

\begin{figure}[h]

\begin{center}
\includegraphics[width=8.cm,height=6.cm]{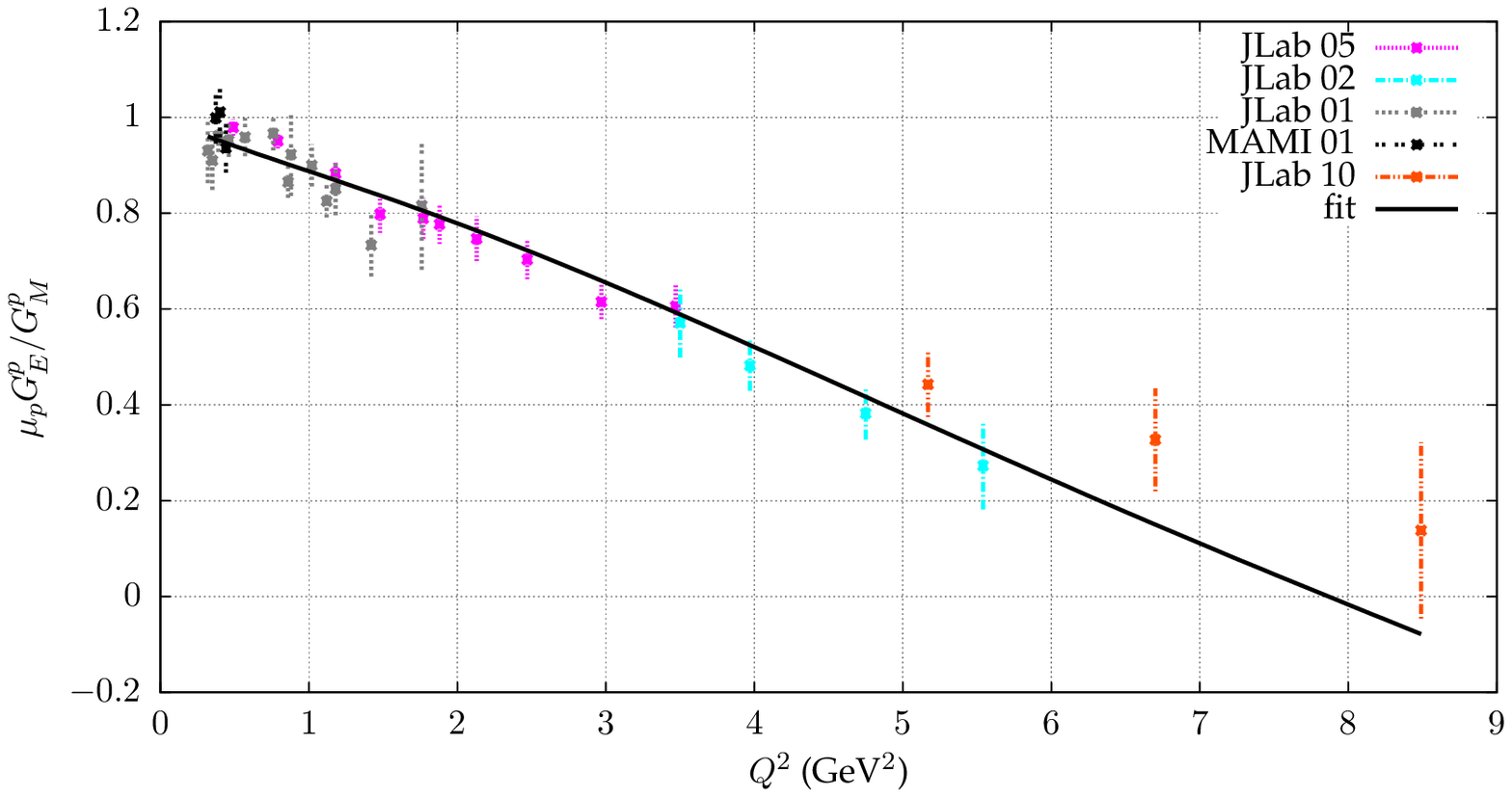}
\includegraphics[width=8.cm,height=6.cm]{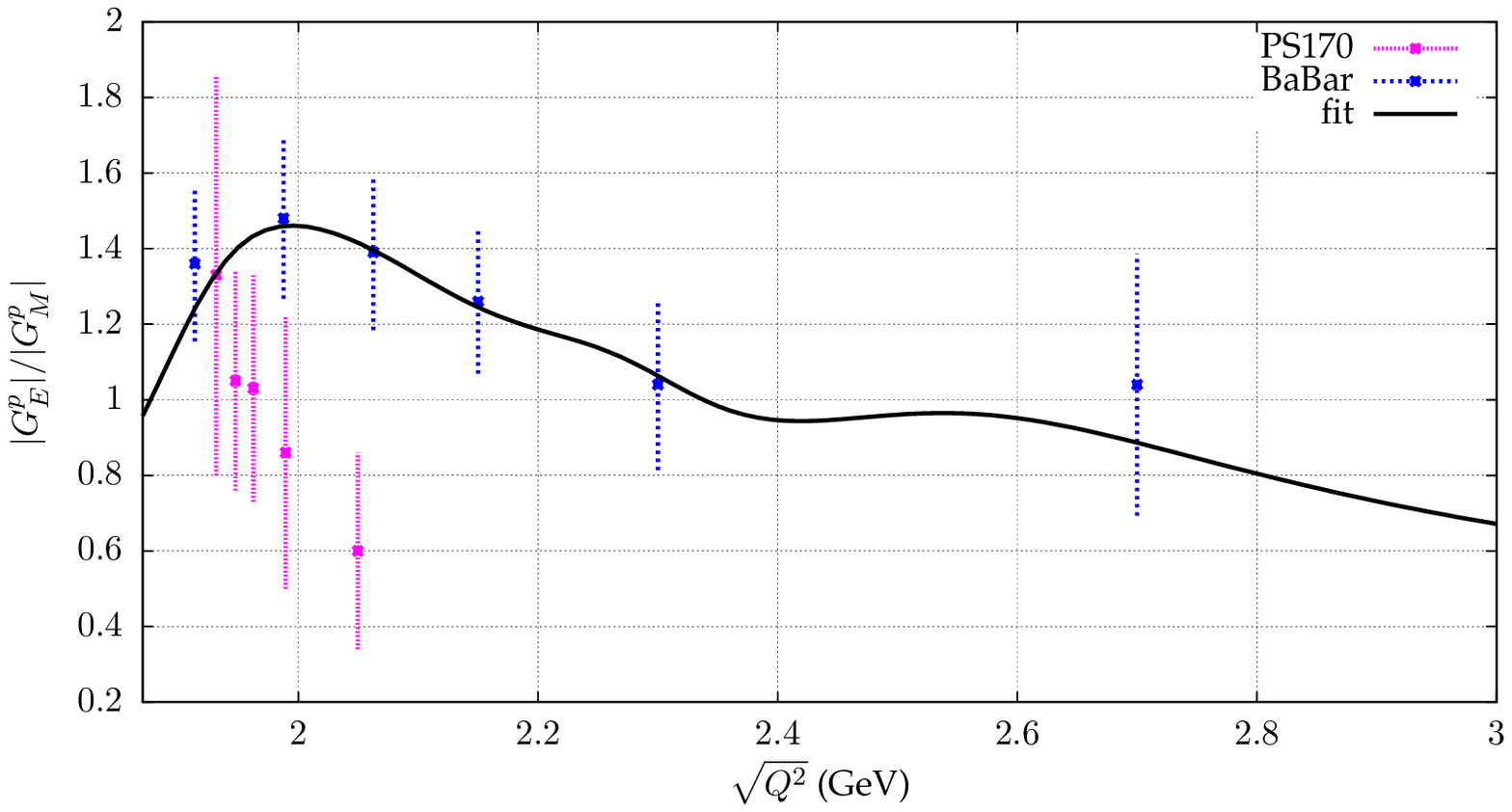}
\caption{(color online)
 \label{pipi4}The experimental data compared to the model fits results.
Ratio of the electric and magnetic proton
 form factors in space-like (upper plot) and time-like (lower plot) regions.
}
\end{center}

\end{figure}

\begin{figure}[h]
 
\begin{center}
\includegraphics[width=8.cm,height=6.cm]{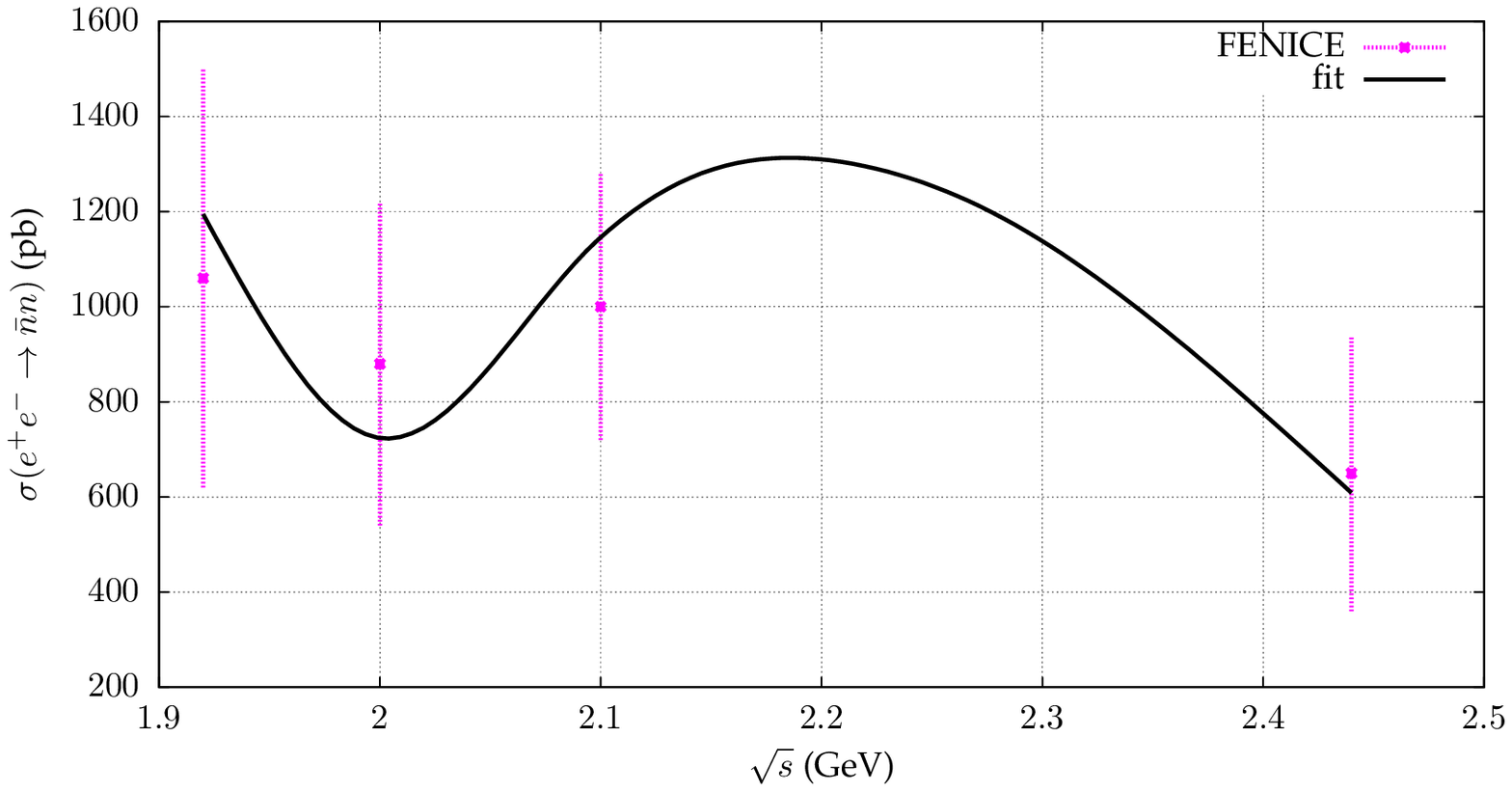}
\includegraphics[width=8.cm,height=6.cm]{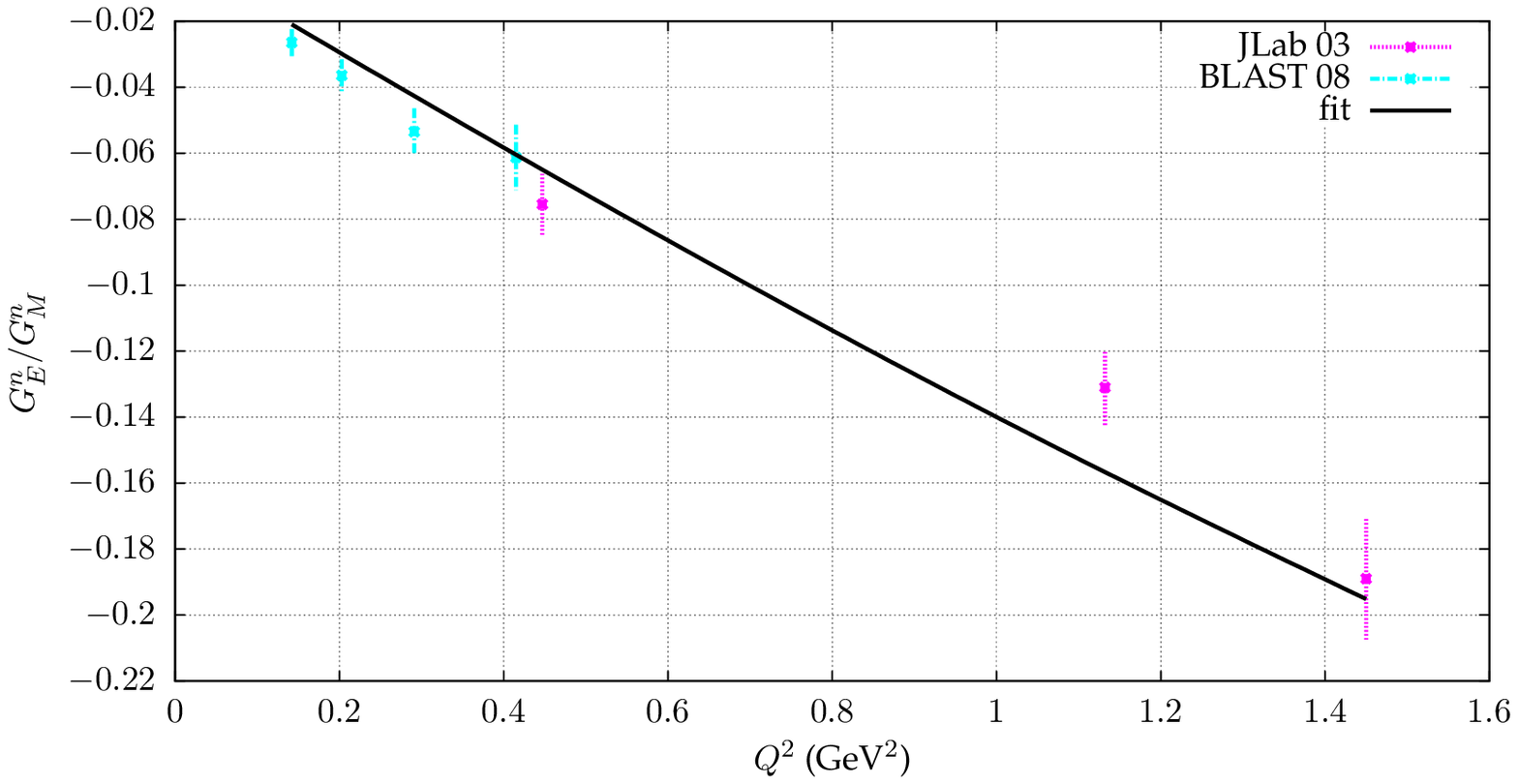}
\caption{(color online)
 \label{pipi5}The experimental data compared to the model fits results.
  Neutron-antineutron production cross section  (upper plot) and ratio
 of the electric and magnetic neutron
 form factors in space-like region (lower plot).
}
\end{center}

\end{figure}
\begin{figure}[h]
 
\begin{center}
\includegraphics[width=8.cm,height=6.cm]{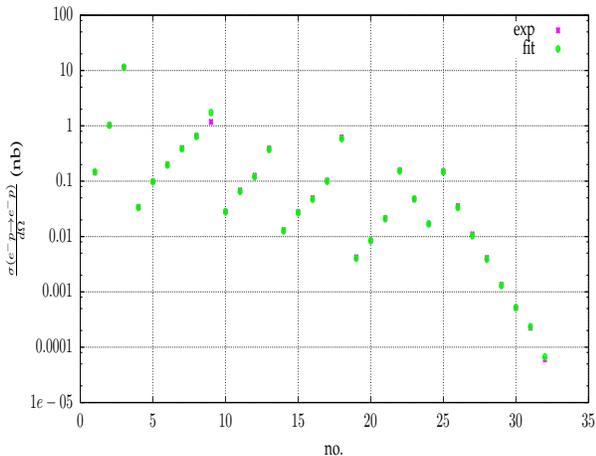}
\caption{(color online)
 \label{splsigma}The experimental data \cite{Andivahis:1994rq} compared
  to the model fit results.
 The data points and fit results are overlaid for most cases.
  On the horizontal axis the entry number from Table IV of 
 \cite{Andivahis:1994rq} is given.
}
\end{center}

\end{figure}

 The masses and widths of the mesons and nucleons were taken
 from PDG \cite{Beringer:1900zz} with the exception of 
  $\rho_{3,4}$ and $\omega_{3,4}$ adopted from kaon form factor model 
 \cite{Czyz:2010hj}
  ($m_{\rho_3} =2.12$~GeV, $\Gamma_{\rho_3} =0.3$~GeV, $m_{\omega_3} =2.0707$~GeV,
  $\Gamma_{\omega_3} =1.03535$~GeV, $m_{\rho_4} =2.32647$~GeV, 
  $\Gamma_{\rho_4} =0.4473$~GeV, $m_{\omega_4} =2.34795$~GeV,
  $\Gamma_{\omega_4} =1.173975$~GeV).
The model parameters, namely the complex $V\bar N N$ couplings,
   were fitted to the following experimental data:
 $e^+e^-\to \bar p p$ cross section \cite{Lees:2013ebn,bes,cleo,dm2,dm22,adone,dm1,fenice},
 $\bar p p \to e^+e^-$ cross section 
 \cite{ps170_1,psrat,Bardin:1991bz,E760,E835_1,E835_2}, $e^+e^-\to \bar n n$  cross section \cite{fenice}, 
ratio of the proton electric and magnetic form factors in the space-like
 \cite{jlab,jlab2002,jlab3,a23} and the time-like \cite{psrat,Lees:2013ebn} 
 regions
 and ratio of the neutron electric and magnetic form factors in the 
 space-like region \cite{E,blast}. 
 The cross sections of the reaction $ep\to ep$, which depend also 
  on the form factors we model, contain also
  non-negligible contributions from two-photon exchange diagrams
  \cite{Carlson:2007sp}. The modelling of these contributions is 
 beyond the scope of this paper and we adopted the following procedure
 to get a reasonable description of this cross section as well:
 We consider only one data set
 \cite{Andivahis:1994rq} covering large range of angles and kinematical
  invariant ($Q^2$). In fit we neglect the contribution from 
  two-photon exchange diagrams and to account for this we enlarge
  the cross section error bars used in the fitting procedure 
  to 10 \% of the cross section. 

 A fit to all the above experimental
 data  leads to unacceptable results ($\chi^2=214$ for 177 data points). 
 The reason is that the  
  PS170 \cite{ps170_1,Bardin:1991bz,psrat}
 and DM2 
 \cite{dm2,dm22}) data 
  are in conflict with the BaBar data \cite{Lees:2013ebn}. 
 It is quite implausible that any model can accommodate
 PS170 and BaBar data sets at the same time 
 as the ratio of the form factors is in evident 
 conflict and,  even more important, both $e^+e^-\to \bar p p$
 and $\bar p p \to e^+e^-$ cross sections are proportional
  to the same combination of the magnetic and electric form factors
  $\left(|G_M^p|^2(1+\cos^2{\theta})+
 \frac{|G_E^p|^2}{\tau}\sin^2{\theta}\right)$ 
 (where $\theta$ is the scattering angle) and thus the same function
 is measured in both cases.

 The model accommodates well the whole data set if one excludes either
 PS170 and DM2 ($\chi^2=124$ for 150 data points) or BaBar  data
 ($\chi^2=107$ for 133 data points). In both cases the  $\chi^2$ values
  are excellent, but each of the models is in  strong
  conflict with the data set which was not fitted. 
  We report here only the details of the fit
 (Tables \ref{tab:fit1} and \ref{tab:pion}, Figs. \ref{pipi}-\ref{pipi5}), were PS170 and DM2 data
  were excluded. Nevertheless it would be highly desirable to confirm the 
  BaBar data by an independent measurement with  similar precision.
    One observes  (Fig. \ref{pipi5}) 
 that for model building a more accurate 
 neutron-anti-neutron cross section would be desirable as 
the presently available data give little constraints on the model 
  parameters.

\section{\label{sec3} FSR corrections to $e^+e^-\to \bar p p \gamma$ }
 
 The BaBar data set was obtained with the radiative return method,
  and final state photon(s) emission was argued to be negligible
 \cite{Lees:2013ebn}.
  As an independent data set might be obtained for example via
 radiative return method with higher accuracy
 the role of final state emission has
 to be reconsidered. As evident
 from the previous section photon-nucleon interactions are not well known.
 Modelling of real photon emission from a proton is thus difficult. 
  We follow here the scheme adopted successfully in \cite{Czyz:2003ue}
  for final state emission from charged pions. For the proton 
  the situation is more complicated, as due to the presence of the Pauli
  form factor at LO, the model is not normalisable.  Here we adopted
  the simplest model assuming that real photon emission from
  a proton (anti-proton) looks like emission from a point-like charged
  particle. This means that it is identical to the emission
 from muons \cite{Czyz:2004rj}.  For the virtual corrections 
one cannot simply adapt the corrections
   from the muon case. Due to the presence of the $F_2$ form factor
   they are not the same and the corrections proportional to $F_1$
   and $F_2$ are not expected to be identical. Moreover as the theory
   based on the interaction Lagrangian ${\cal L} = A^\mu J_\mu$
   with $A_\mu$ being the electromagnetic field and $J_\mu$ defined
   in Eq.(\ref{pradh}), is not renormalisable, further complications arise.
   They will not be addressed in this paper.
  For the virtual corrections 
 we have used an overall factor, multiplying zero-, one- and two-photon emission parts:

\begin{equation}
C(Q^2)=f(\pi\alpha/\beta)-f(\pi\alpha)+1,
\label{cfh}
\end{equation}
where
 \begin{equation}
f(x)=\frac{x}{1-\exp{(-x)}} \ , \ \ \ \ \beta =\sqrt{1-4m_p^2/Q^2} \ ,
\end{equation}
 $m_p$ is the proton mass and $Q^2$ is the invariant mass
  of proton-anti-proton pair.

 At small proton velocities $C(Q^2)$ reproduces
 the usual Coulomb factor which resumes the leading radiative corrections
 for small velocities,
  while at large invariant masses  $C(Q^2\to \infty) \to 1 $. 
 In addition a correction of the form

\begin{equation}
\Delta_{final}=\frac{2\alpha}{\pi}\left[\frac{(1+\beta^2)}{2\beta} \log{\frac{Q^2(1+\beta)^2}{4m_p^2}}-1\right]\log{2w} 
\end{equation}
 with $w=E_{\gamma,min}/\sqrt{s}$ was added , where $E_{\gamma,min}$ 
 is a separation parameter
  between soft and hard parts of the photon phase space. 
 It compensates the divergences arising from integration of a real
 emitted photon with energy $E_\gamma>E_{\gamma,min}$. 
Technically it leads to the replacement
   of
   $\frac{\alpha}{\pi} \eta^{V+S}$ used for muons in \cite{Czyz:2004rj}
  with $\Delta_{final}$.
 The  factor $C$ takes care of the proper threshold behaviour.  
 Both $C$ and $\Delta_{final}$ factors are generic for any model,
  which assumes that for soft photon emission
   proton behaves like a point-like particle.
 In any more elaborated model for the virtual corrections
 there will be additional finite corrections 
 proportional to $\alpha/\pi$, which will depend on the model
 details. In our ansatz we put them to zero. They are not expected to be big
 and their size can be tested using charge asymmetries as proposed in
 \cite{Binner:1999bt,Czyz:2003ue,Czyz:2004nq} for pion pair production. 
 In our opinion without experimental tests of the FSR corrections
  better modelling of these contributions is not possible.

\begin{figure}[h]
\includegraphics[width=8.cm,height=6.cm]{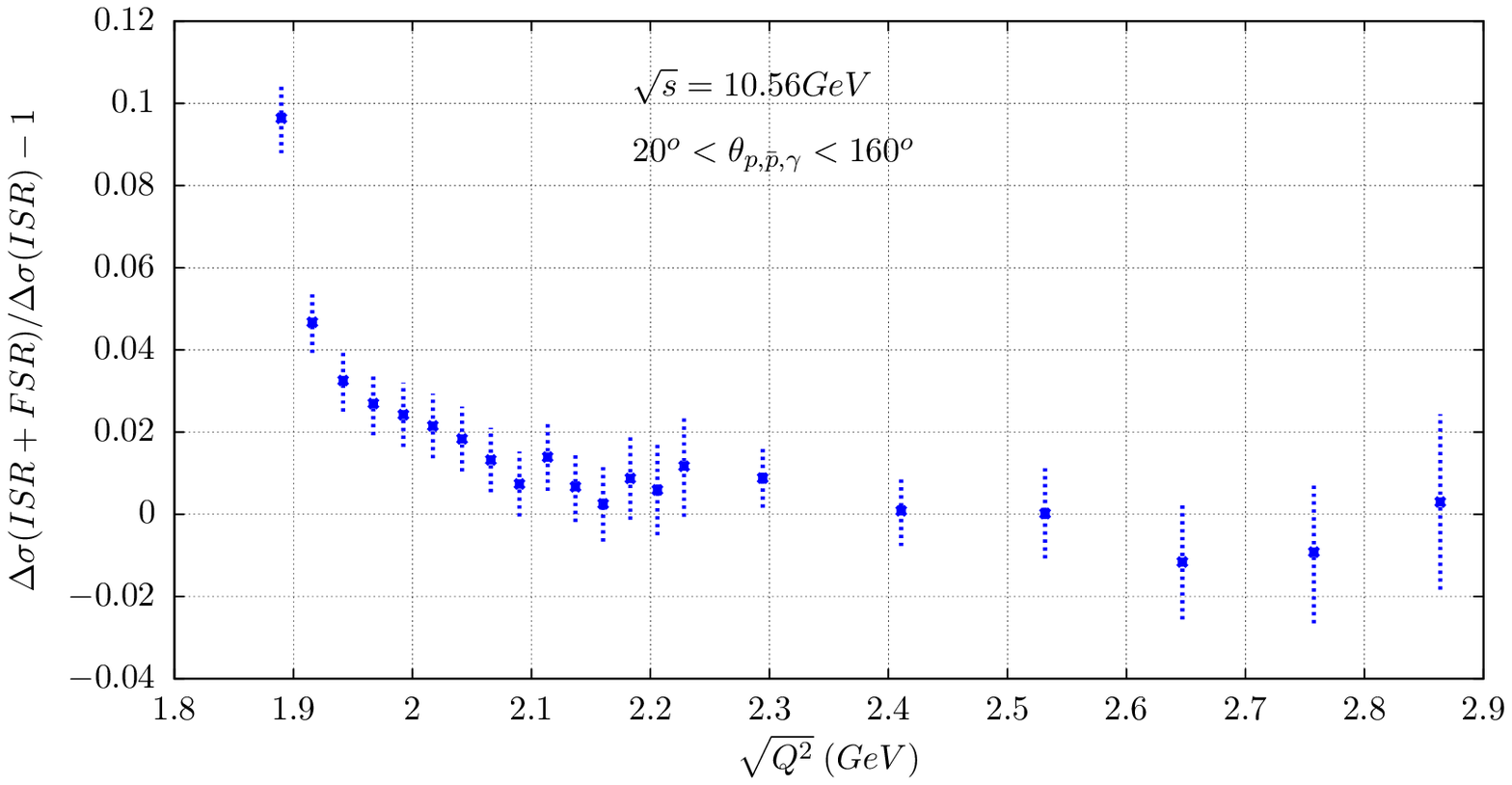}
\includegraphics[width=8.cm,height=6.cm]{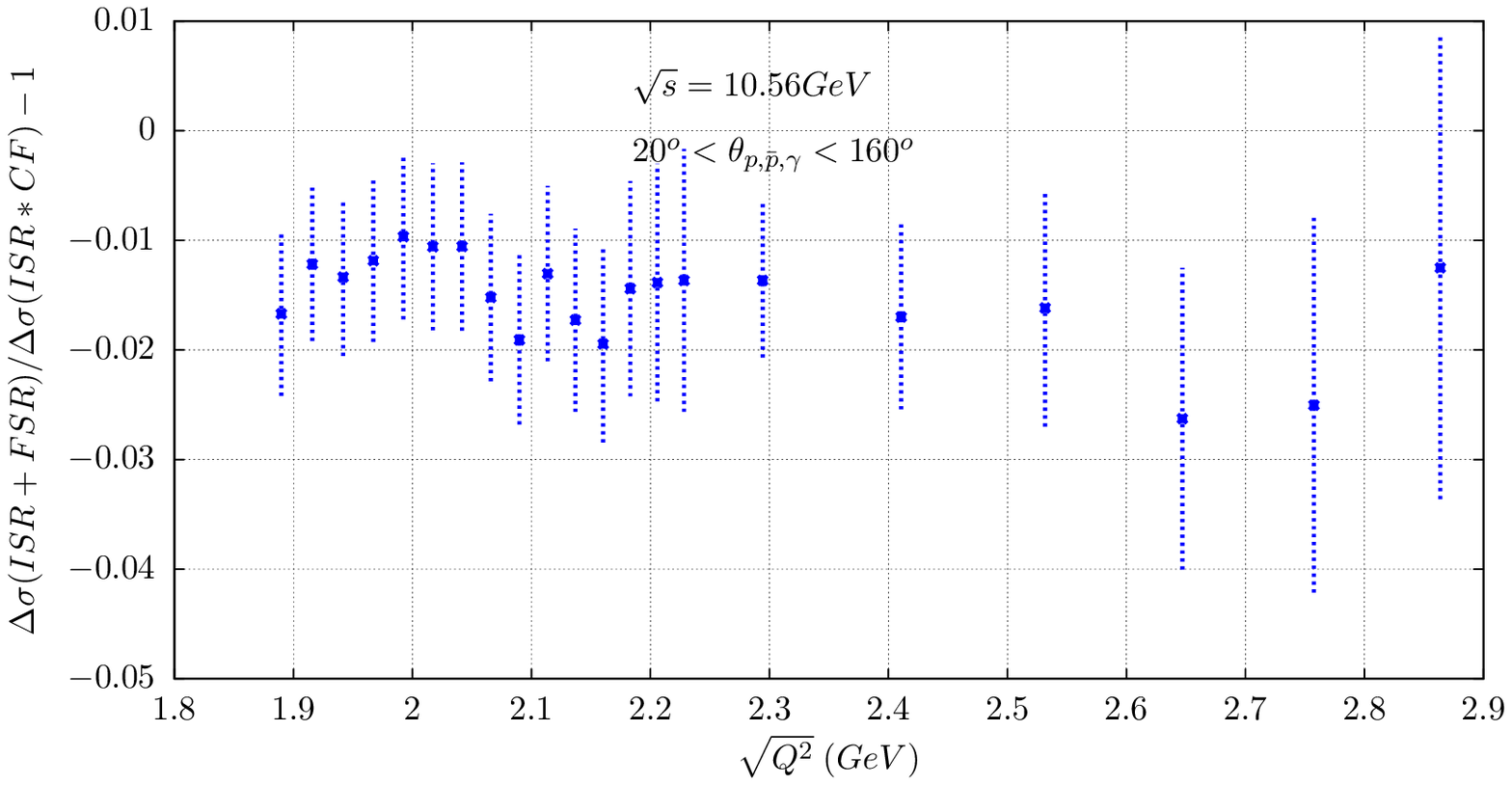}
\caption{(color online)
 \label{ry2}  Relative difference between $ Q^2$ distributions
  calculated at NLO 
  with and without FSR radiative corrections and 
 between complete FSR corrections and FSR corrections where only
  Coulomb factor is included.
}
\end{figure}
\begin{figure}[h]
\includegraphics[width=8.cm,height=6.cm]{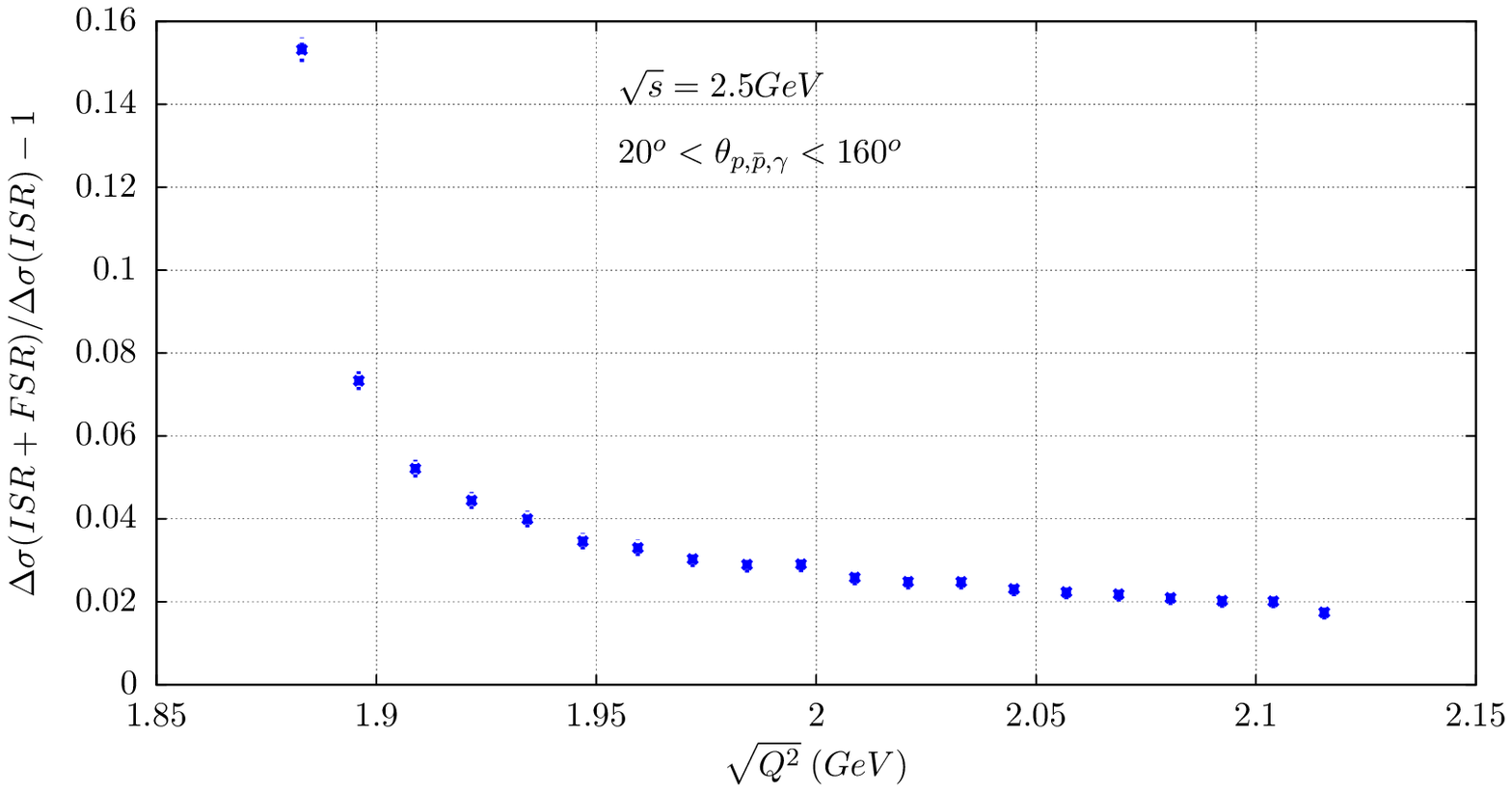}
\includegraphics[width=8.cm,height=6.cm]{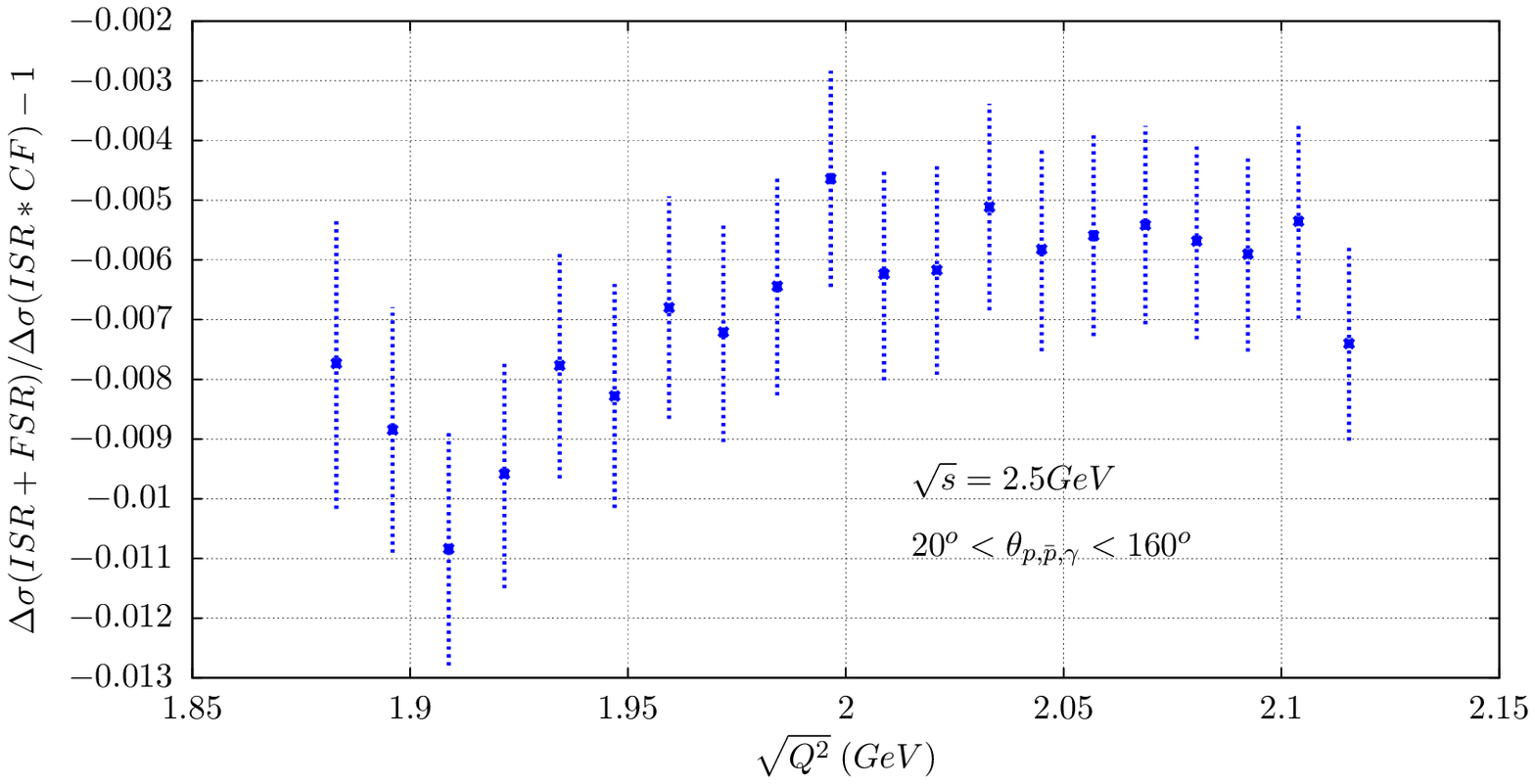}
\caption{(color online)
 \label{ry3} Relative difference between $ Q^2$ distributions
  calculated at NLO 
  with and without FSR radiative corrections and 
 between complete FSR corrections and FSR corrections where only
  Coulomb factor is included. 
}
\end{figure}

  The size of FSR radiative corrections 
  depends both on energy of an experiment and
  on the event selection used and  can be both negative or positive.
  In Fig. \ref{ry2} we show its size for an event selection close to the 
  one used by BaBar. We show there a relative difference of the ISR cross
 section calculated at NLO and the ISR cross section corrected with
 with Coulomb factor
 \begin{equation}
   C_{F}(Q^2)=f(\pi\alpha/\beta) .
\label{cfht}
\end{equation}

 The difference between the FSR correction calculated 
 with Coulomb factor only and FSR at NLO is also shown in Fig. \ref{ry2}.
 Its typical size is of order one percent. 
In  Fig. \ref{ry3} we show the same differences calculated
 at a possible BES-III energy. The FSR corrections not included
  in the Coulomb factor $C_{F}(Q^2)$ are here even lower then at
 BaBar energy, but still of order of one percent at low 
 proton-anti-proton pair
 invariant masses.
  As the FSR corrections for proton antiproton pair production were not
  tested experimentally 
 the one percent contribution coming from our model should
  be taken conservatively as an estimate of the accuracy of the
 modelling of the FSR. 
\begin{figure}[ht]
\vskip 0.2cm
\includegraphics[width=8.cm,height=6.cm]{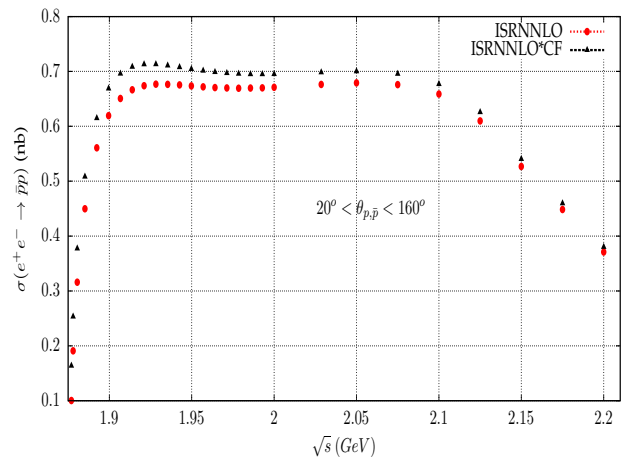}
\caption{(color online)
 \label{ry4} The cross section of the reaction $e^+e^-\to\bar p p$ with
ISR at NNLO corrections included is compared to the same cross section
 where in addition FSR Coulomb corrections were added. 
}
\end{figure}

 FSR corrections were also implemented for proton antiproton final
 state in the 'scan' mode of the PHOKHARA Monte Carlo event
 generator \cite{Czyz:2013xga}.
 Only the Coulomb factor Eq. (\ref{cfht}) was taken into account. Its 
 size is demonstrated in Fig. \ref{ry4}. From this figure
 one can see that in any scan experiment in the region close to the
 threshold where in principle one can test the resummation of
 radiative corrections the beam energy smearing effects 
 (where  beam spread is typically 1-2 MeV) will obscure the effect.
 The distance between the first two points is taken as 1.5 MeV while
  between the 
 second and the third point the step size is 2 MeV.

 The event generator PHOKHARA 9.1, which includes the new nucleon 
 form factors and final state corrections is available at the 
 web page {\tt http://ific.uv.es/$\sim$rodrigo/phokhara/ }.
  The newly added part of the code proportional to the Pauli form factor
 with real photon emission was tested using an independently 
 written code.  In the distributed code 
 the helicity amplitude method is used. Tests on independence of the
 cross section on the soft-hard photon separation parameter $w$ were
 also performed with the accuracy not worse than 0.03\%.

\section{\label{sec4} Conclusions}

We have constructed nucleon form factors on the basis of generalised vector dominance which are consistent with recent BaBar results for
proton-antiproton production through the radiative return, with older data for neutron-antineutron production and with results for electron-nucleon scattering. Furthermore these form factors exhibit the the high energy behaviour predicted from perturbative QCD. We have considered the effect of final state radiation, demonstrating its smallness for typical experimental cuts. The new form factors and real as well as virtual final state radiation have been implemented in the Monte Carlo generator PHOKHARA.

\begin{acknowledgments}
Henryk Czy\.z is grateful for the support and the kind hospitality 
of the Institut f{\"u}r Theoretische Teilchenphysik
 of the Karlsruhe Institute of Technology. 
\end{acknowledgments}

\bibliography{biblio}

\end{document}